# Demonstration of novel high-power acoustic through-the-wall sensor

Franklin Felber*
Starmark, Inc., P. O. Box 270710, San Diego, CA 92198

## ABSTRACT

A high-power acoustic sensor, capable of detecting and tracking persons through steel walls of cargo containers, trailer truck bodies, and train cars, has been developed and demonstrated.  The sensor is based on a new concept for narrowband mechanical-impact acoustic transmitters and matched resonant receivers.  The lightweight, compact, and low-cost transmitters produce high-power acoustic pulses at one or more discrete frequencies with little input power.  The energy for each pulse is accumulated over long times at low powers, like a mousetrap, and therefore can be operated with ordinary batteries and no power conditioning.  A breadboard impact-transmitter and matched-receiver system that detected human motion through thick walls with only rudimentary signal processing is described, and results are presented.  A conceptual design is presented of an acoustic through-the-wall sensor, costing about $10,000 per unit and capable of remotely and non-intrusively scanning steel cargo containers for stowaways at a rate of two containers per minute.  Advantages of acoustic through-the-wall sensors over radar are: Sound penetrates metal walls; and acoustic sensors are sensitive to small and slow motions, and so can detect stationary persons by breathing motion alone.  Other attractive features include: high-resolution locating and tracking; portability; low cost; quick and easy preparation and deployment; and near-real-time data processing and display.  These features provide a robust stand-alone through-the-wall surveillance capability or an excellent complement to a radar sensor.

**Keywords:** Through the wall, acoustic surveillance, acoustic detection, acoustic sensors, mechanical impact transducers, mechanical acoustic transmitters, personnel detection, stowaway detection

## 1. INTRODUCTION

This paper presents a new concept for mechanical transducers that are particularly well suited for efficiently and inexpensively producing and coupling high-power acoustic pulses into dense media, like walls.  The transducer system advances reported in this paper are:
(i) a new mechanical transmitter that is compact, lightweight, low-cost, and can operate on battery power, yet produces acoustic pulses in dense media at one or more frequencies that are orders of magnitude more powerful than those produced by alternative transmitters of comparable scale;
(ii) a resonant receiver matched to the transmitter and designed for efficient coupling to dense media and for enhancing with multiple sensors the signal-to-noise ratio (S/N) of the received signal before signal processing;
(iii) effective configurations for efficiently coupling the transmitter and matched receiver to dense media; and
(iv) methods of signal processing that increase S/N by several additional tens of dB.

Sections 2 through 4 present the innovative concepts of mechanical impact transmitters and matched resonant receivers and their effective coupling into walls.  Sections 5 through 7 focus on the application of detecting humans through walls, particularly the steel walls of cargo containers, trailer trucks, and train cars, which cannot be penetrated by radar.

An acoustic means of through-wall surveillance (TWS) and tracking was discovered by the author in 1997.  While conducting programs to develop acoustic sensors of concealed weapons[1–6], it was found that commercial off-the-shelf (COTS) ultrasound transducers placed against a solid barrier produced an echo from the other side of the barrier that changed when someone moved behind the barrier.  Then by subtracting successive echo pulse waveforms, the difference waveform, through destructive interference, revealed only those persons or objects that moved between pulses; the echo pulse waveforms returned from stationary objects canceled each other.  The round-trip time of each pulse returned to a receiver from a moving person or object indicated its range, and triangulating the ranges to multiple receivers indicated its location.

*felber@san.rr.com; phone 1 858 676-0055; fax 1 858 676-0003

The primary alternatives to acoustic TWS since the mid-1990s have been radar-microwave and passive millimeter-wave sensors[7–20]. Passive millimeter-wave sensors required illumination of the targets by millimeter-wave radiation from the sky, which effectively limited their applicability to finding persons in areas open to the sky[11,14]. Radar-microwave sensors, like the Hughes Motion Detection Radar[10,12,16], differential radar[13], radar 'flashlight'[15], Time Domain's RadarVision and SoldierVision[17–19], and Livermore's Urban Eyes[20], were limited by attenuation in walls to long microwave wavelengths, typically S band and longer, which did not allow detection of the millimeter-scale motions of 'motionless' people. More significantly, radar could not penetrate metal or metal-lined walls or even the aluminum-backed fiberglass insulation typically found in homes and buildings.

SAIC's Vehicle and Cargo Inspection System[21] (VACIS®) is designed to penetrate 15 cm of steel and image the entire contents of cargo containers and trailer trucks, but the system components must be big enough to span the trucks and cargo containers it scans. Also, highly ionizing gamma radiation is not allowed for use on humans in TWS applications. In conjunction with acoustic TWS, however, harmful gamma-ray doses to stowaways[22] by VACIS® can be avoided by prescreening cargo containers with an acoustic sensor.

The foremost advantage of acoustic TWS is that sound penetrates metal walls almost as well as all other solid wall materials, and does so with harmless non-ionizing radiation. Another significant advantage is that acoustic TWS is sensitive to motions smaller than 1/10 of a wavelength and can potentially find stationary persons by the breathing motion of their chest cavities alone.

Following the discovery of acoustic TWS, we developed an acoustic TWS breadboard system[23], shown schematically in Fig. 1(a). The system operated by transmitting narrowband acoustic pulses through a wall and receiving the reflected signals from all objects inside the room at two laterally separated receivers. Signal processing algorithms were used to track multiple persons through a wall and to detect even the millimeter-scale chest motion of a person lying down and breathing quietly. Figure 1(b) shows an implementation of our tracking algorithm on TWS binaural waveforms from the breadboard TWS monitor of Fig. 1(a).

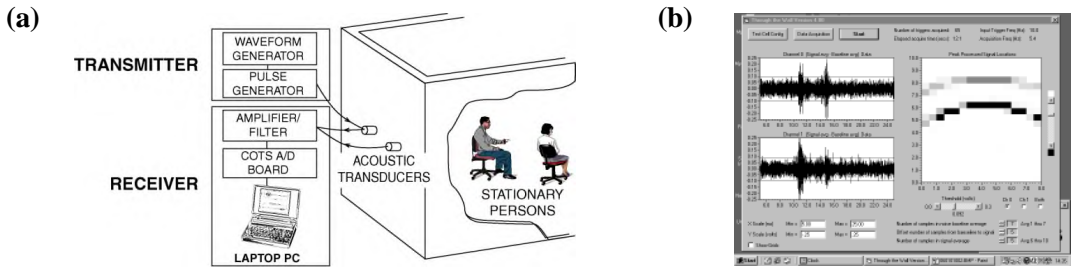

Figure 1. (a) Schematic of early breadboard acoustic TWS monitor based on COTS transducers. (b) Monitor shows two stationary persons sitting 6 ft and 8 ft behind a 4.3-cm solid wooden door.

Although the COTS ceramic piezoelectric transducers used in this system were operated near their damage threshold, 3200 volts peak-to-peak, TWS performance was severely power-limited, especially since *these transmitters lose their resonant qualities when placed against a wall*. When placed against a wall, their quality factor $Q$ is reduced from 25 by more than an order of magnitude, and the acoustic power, which scales as $Q^2$, falls by a factor of hundreds. Not even COTS underwater transducers[24] could provide the high-power, narrowband acoustic pulses needed for TWS. As seen in Fig. 1(b), the S/N was marginal at a range less than 3 m through a mere 4.3-cm-thick wooden door.

In response to the need for high-power TWS transducers, we developed a narrowband, tunable mechanical transmitter for the Marine Corps Systems Command (MCSC) that was capable of scaling up the transmit/receive power of the acoustic TWS system by more than 72 dB (a factor of 16 million), but that was lighter than the COTS transmitter it replaced[25]. Another significant advantage of this mechanical "MCSC transmitter" was safety from high voltages, because it operated on a single 9-V battery.

The nontunable, resonant transmitter concepts in Sec. 2 of this paper have important modifications of the MCSC transmitter design that boost peak power output by orders of magnitude. The high-power transmitter concepts were developed for both underwater and through-wall applications[26]. Since issues concerning design of transducers and coupling to walls are related to those issues for all dense media, the TWS transducer technology presented in Secs. 3 and 4 is applicable, with appropriate modifications, to production and coupling of acoustic pulses in all liquid and solid media as well. The coupling into solid media is discussed in Sec. 4. The advantages of the mechanical transducers described in this paper – high acoustic pulse power, low electrical power, high efficiency, light weight, compact size, low cost, low-voltage long-duration battery operation – are as attractive for underwater systems as for TWS.

In short, COTS transmitters and power amplifiers, even those that are inefficient and have relatively less demanding electrical power requirements, are expensive. The power conditioning subsystems generally are heavy and bulky and require the order of kilovolt voltages. None approaches a cost of hundreds of dollars per unit, which is a reasonable goal for battery-operated, amplifier-free, high-power impact transmitters.

## 2. CONCEPT OF MECHANICAL IMPACT TRANSMITTERS

We discovered[27] a highly effective and efficient way to couple acoustic energy into a dense medium, such as a wall or water: *A properly designed thin plate transduces a <u>mechanical impulse</u> to acoustic energy in a dense medium with high efficiency at the resonant mode frequencies of the plate.*

The concept for transduction by impact excitation of a thin plate is shown in Fig. 2(a). An electromechanical actuator propels a mass, called an impactor, at a thin plate. The mass impacts the plate and bounces back, exciting the fundamental mode if striking the center of a symmetrical plate, and exciting higher frequency modes if striking off center. The thin plate then rings down, delivering much of its kinetic energy to acoustic radiation in the medium, if coupled properly.

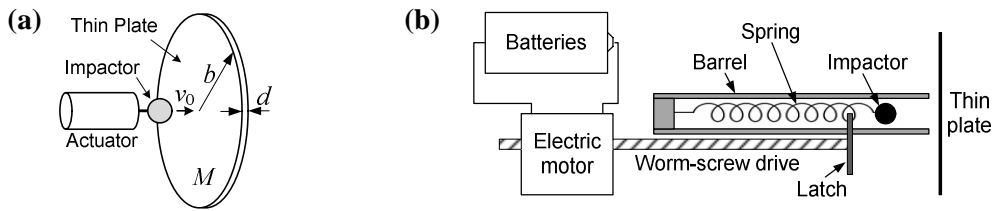

Figure 2. Conceptual sketches of: (a) impact transmitter; and (b) electromechanical pull-type actuator.

The 'impact transmitter' performs much like a spring-loaded mousetrap. Potential energy is accumulated and stored at low power over a relatively long time. Then by opening a latch or some other mechanical release mechanism, the potential energy is suddenly transformed to kinetic energy and delivered to a target in an instant at high power.

Although impact transmitters can have high energy efficiency, that is not their principal advantage or what will allow the transmitters to be manufactured at low cost. *The principal advantage of impact transducers is that the energy for each pulse is accumulated over long times at low powers, and therefore can be operated with ordinary batteries and no power conditioning.* For example, the MCSC transmitter, driven by a small electric motor powered by one 9-volt transistor battery, produced orders of magnitude higher acoustic power than the breadboard TWS system in Fig. 1, driven by thousands of volts peak-to-peak.

Figure 3 shows several means of mechanically driving a thin plate, which is part of the interface of the transmitter with the dense medium. Each employs efficient electric motors and/or actuators with decades of development and commercial and industrial use.

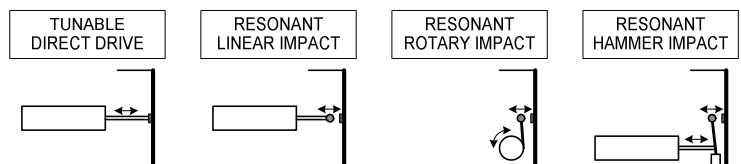

Figure 3. Mechanical means of driving a thin-plate transmitter with motors and/or actuators.

The direct-drive mechanism, familiar from uses in high-power hydraulic transmitters and loudspeakers, produces a displacement of the plate limited to the stroke of the motor. The other three mechanisms shown in Fig. 3 are all impact-type transmitters. The pull-type actuator in Fig. 2(b) is an example of a linear-impact mechanical actuator. The rotary-impact and hammer-impact transmitters employ mechanical advantage.

General design equations for an impact transmitter follow. For specificity, a circular plate with edge-clamped boundary conditions is chosen, but the design equations are easily modified to other plate shapes and boundary conditions.

The normal displacement $z$ at radius $r$ and time $t$ of a thin circular plate, rigidly clamped all around its circumference, vibrating in its fundamental mode, is[28]

$$z(r,t) = z_0(t)\sin(\omega t)[J_0(3.2r/b) + 0.0555 I_0(3.2r/b)]/1.0555 . \tag{1}$$

Here, $z_0(t)$ is the displacement amplitude at the center of the plate, which decays with time; $J_0$ and $I_0$ are the zero-order Bessel function and modified Bessel function of the first kind, respectively; $b$ is the radius of the plate (to the clamped edge); and $\omega$ is the angular frequency of oscillation. The frequency of the fundamental mode of the thin plate in air is[28]

$$f_p = 0.47(d/b^2)[Y/\rho_p(1-\sigma^2)]^{1/2} , \tag{2}$$

where $d$ is the thickness, $\rho_p$ is the density, $Y$ is the Young's modulus, and $\sigma$ is the Poisson's ratio of the plate material.

From Eq. (1) and its first derivative with respect to time, the maximum kinetic energy of the whole plate is calculated as a function of the maximum plate velocity at the center, $u_0$, and is found to be $0.182(Mu_0^2/2)$, which is equal to the maximum kinetic energy of a flat circular piston having the same velocity, $u_0$, but a mass of only $0.182M$. Because only the central region of the edge-clamped plate moves much, the active plate mass $M_a$ is only 18.2% of the actual mass $M$. That means the active area of the plate is $A_a = 0.182A$. Different boundary conditions would result in different active masses and areas.

The total specific energy of the plate decays exponentially as

$$E(t) = E_0 \exp(-\Gamma t) , \tag{3}$$

where $\Gamma$ is the decay rate, and $E_0$ is the initial total specific energy after the impactor has imparted its energy. The quality factor of the oscillator is

$$Q = \omega/\Gamma . \tag{4}$$

By definition[24], the transmitter bandwidth is $B = \omega/2\pi Q$, or $B = \Gamma/2\pi$. The potential energy of the active mass is of the form $K(z_0 \sin \omega_0 t)^2/2$, where the effective spring constant of the plate is

$$K = M_a \omega^2 = M_a (2\pi f_p)^2 . \tag{5}$$

The potential energy is a maximum when the plate reaches its maximum displacement amplitude, $z_0(t)$, at which point the potential energy $Kz_0^2(t)/2$ equals the total energy $M_a E(t)$. The displacement of the plate on axis, therefore, is

$$z(t) = (2E_0/\omega^2)^{1/2} \sin(\omega t)\exp(-\Gamma t/2) , \tag{6}$$

and the velocity of the plate on axis, for $|d\ln(u)/dt| \ll \omega$, is

$$u(t) = (2E_0)^{1/2} \cos(\omega t)\exp(-\Gamma t/2) . \tag{7}$$

The average rate of energy loss from the plate, some fraction (calculated in Sec. 4) of which is coupled into the wall, is

$$P(t) = \Gamma M_a E(t) \,. \tag{8}$$

In general, an efficient transmitter uses a lightweight plate material with low density and high sound speed, such as aluminum, which has the following properties[28]: $\rho_p = 2700\,\text{kg/m}^3$, $Y = 7.1 \times 10^{10}\,\text{Pa}$, $\sigma = 0.33$. The optimal impactor for a plate following these design equations is found by trading off limitations imposed by optimizing efficiency with respect to: (i) frequency; (ii) power; and (iii) duration of impact. The choice of impactor material is governed by the need to have the impactor rebound from the plate quickly. The duration of the impact on the plate should be a small fraction of the oscillation period of the plate, in order that the impactor not impede the first oscillation of the plate after impact. Small, hard impactors produce impacts of short duration. The impact duration is proportional to the diameter of a spherical impactor[29], and hard impactors produce more nearly elastic collisions. A good choice of impactor material, therefore, is steel.

## 3. CONCEPT OF MATCHED RESONANT RECEIVERS

As a general rule, virtually any type of acoustic transmitter, with appropriate modifications, can also be operated as a receiver. The impact transmitter is no exception. The thin plate that is the active surface in an impact transmitter can also be used to receive returning echo signals in TWS or sonar applications.

The receiver plate can be the same plate used for transmission, but it does not have to be. For many applications, it may be preferable to have the receiver plate be separate from the transmitter plate, even if both plates are collocated. A common reason for wanting to have separate transmitters and receivers is that the echo signal is expected to return to the receiver before the oscillations of the transmitter plate have decayed sufficiently. In such cases, the transmitter and receiver plates might not only need to be separate, but also acoustically isolated from each other. For related reasons, the receiver might also need to be time-gated, so that it does not start 'listening' for returning echo signals until some number of milliseconds after the transmitter generates a pulse. For the reasons that follow, however, if the transmitter and receiver plates are separate, they may need to be identical, particularly for narrowband applications.

The impact transmitter produces sound predominantly at the frequencies of the modes of vibration of the plate that are excited by the impact. The 'natural' resonant-mode frequencies of a plate are functions of the material, thickness, size, and shape of the plate, but also of how the plate is mounted on the transmitter body. The normal mode frequencies depend on the boundary conditions of the plate, such as at what edges the plate is mounted and whether the rim is rigidly edge-clamped or just fixed in place. Where the impactor strikes the plate will determine the linear combination of normal modes that will be excited.

Figure 4 shows the complete breadboard TWS transducer system. From left to right and back to front, the items in Fig. 4 are: Impact transmitter; resonant receiver plate assembly (rear face); triggering cable; 2 piezo-film sensors; and a COTS filter/amplifier with transformer and power cord.

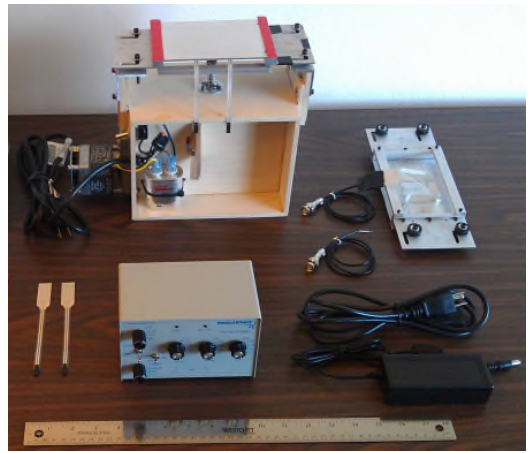

Figure 4. Complete high-power mechanical breadboard TWS system, size relative to 18-in ruler.

Particularly in TWS applications, in which there is a premium on narrowband pulses to enhance S/N, it is important that the receiver plate be identical to the transmitter plate and that their mountings also be identical. The breadboard TWS impact-transducer system, shown in Fig. 4, has a bandwidth of 7 Hz at a resonant frequency of 2186 Hz, for a fractional bandwidth of 0.003. The plates and their mountings need to be identical to a degree of accuracy such that the bandwidths of the useful resonances at the receiver substantially overlap the corresponding bandwidths at the transmitter. Since the receiver plate differs from the transmitter plate in that it does not need to be attached to a more massive structure like the transmitter, as shown in Fig. 4, the transmitter plate should be acoustically isolated from the transmitter structure, so that the resonances of the transmitter structure do not couple to and affect the resonances of the transmitter plate. This isolation was accomplished, as described in Sec. 6, by rubber acoustic isolators.

The preferred sensors for the receivers, whether in-air TWS sensors or underwater, are piezoelectric film sensors with attached leads and adhesive backing like transparent tape. The type of piezo-film sensors used in the breadboard TWS system described in Sec. 6 are adhesive dart sensors with flexible lead attachments. The piezo-film sensors are affixed to surfaces where the stress is greatest for the vibrational mode to be measured.

In general, a multiplicity of piezo-film sensors can be and should be attached to the (inside) surface of the thin plate of the receiver. The sensors should be attached with their polarities, corresponding to the vibrational mode that is being measured, in series. If $n$ sensors attached in series each measure the same voltage response with the same polarity, the signal strength will be amplified by $n^2$, because signal strength scales as voltage squared. For example, the receiver plate in Fig. 4 has four piezo-film sensors mounted on its rear face in a square array around the center of the plate, enhancing the signal voltage by a factor of 4 over a single sensor and the signal strength by 12 dB.

For some vibrational modes, such as dipole modes, one piezo-film sensor may be stretched over a convex bend while another is compressed over a concave bend. In that case, their voltage polarities will be opposite each other, and their leads should be connected positive to positive or negative to negative in order for their voltage signals to add constructively.

## 4. EFFICIENT COUPLING OF TRANSDUCERS TO WALLS

Developing high-power transmitters and sensitive receivers is only part of the challenge for acoustic TWS. The transducers must not only be designed for efficient coupling of their acoustic energy to solid walls, but the transducer/wall interface itself is of critical importance to overall coupling efficiency. Indeed, others who have tried to develop acoustic TWS systems to compensate for the shortcomings of radar have found that "the interface between the acoustic transducers and the surface of the wall" was a challenge that "needed to be overcome"[19]. This section discloses preferred methods of coupling mechanical impact transducers to walls.

The method used to couple a transmitter to a wall can make a tremendous difference in performance, increasing the signal strength by orders of magnitude. With good coupling, the transmitter turns the wall to which it is coupled into a soundboard, like the soundboard of a piano, that vibrates at the same frequencies as the thin plate, but produces a louder sound because it moves a much greater volume of air. A good coupling of a transmitter to a wall amplifies the sound the same way holding a tuning fork to a tabletop amplifies the sound.

In a string instrument, like a piano or guitar, the strings vibrate against the soundboard usually via some sort of bridge. An effective 'bridge' to couple both the thin plates of the transmitter and the receiver to the wall was found to be two strips of 1-mm-thick, heavy-duty mounting tape fastened to two ends of the thin plate, seen at the ends of the transmitter plate in Fig. 4. The mounting tape was applied over the edge-clamped ends of the transmitter and receiver plates, and not over the active oscillating area, so that the tape would not affect the resonances when pressed or stuck to a wall.

Consider the transmission of a normally-incident plane sound wave from a solid plate into a solid wall across an air gap of width $L$. The intensity transmission coefficient from the plate into the wall is[28]

$$T_I = 4\left[ 2 + \left(\frac{Z_W}{Z_P} + \frac{Z_P}{Z_W}\right)\cos^2 kL + \left(\frac{Z_A^{\ 2}}{Z_P Z_W} + \frac{Z_P Z_W}{Z_A^{\ 2}}\right)\sin^2 kL \right]^{-1} , \qquad (9)$$

conservatively neglecting forward reflections from the back surface of the plate. Here, $Z_P$, $Z_A$, and $Z_W$ are the characteristic acoustic impedances of the plate, air gap, and wall, respectively, and $k$ is the wavenumber of the sound in air. Regardless of the plate and wall materials, since $Z_A \ll Z_P$ and $Z_A \ll Z_W$, the intensity transmission coefficient is

$$T_I = (2/kL)^2 Z_A^2 / Z_P Z_W, \qquad (10)$$

for air gaps much narrower than a wavelength ($kL \ll 1$).

This scaling of transmission with air-gap thickness, $T_I \sim L^{-2}$, is to be expected. The difference between an air gap of 1 mm and 0.2 mm is 14 dB in transmission into a wall. During operation of the transmitter, coupling of the transmitter plate to the wall seemed to be improved by applying 5 to 10 pounds of force, probably because the force squeezed the foam tape and the air gap to a thickness much less than the original 1-mm thickness. The air gap $L$ should not be made so thin, though, that the maximum amplitude of the plate vibrations $z_0$ causes the plate to contact the wall. That is, transmission into the wall is maximized for the narrowest gap subject to $L > z_0$.

## 5. SIGNAL PROCESSING FOR THROUGH-WALL SURVEILLANCE (TWS)

A TWS sensor, such as a radar sensor, that cannot detect stationary persons may be worse in some circumstances than no TWS sensor at all. Such a sensor can easily give the user a false confidence that he has accounted for all persons on the other side of a wall, which could make a life-or-death difference in certain military operations in urban terrain, law-enforcement, or first-responder situations. Particularly in scenarios involving clearing facilities, or in which dismounted infantry or law-enforcement personnel must enter facilities that may harbor militants or armed and dangerous persons, lives could be saved by the ability to detect people who are unconscious, sleeping, tightly bound, or otherwise immobile. Lives could even be saved of those trying to escape detection by remaining motionless, like stowaways inside cargo containers, who are at risk of death by dehydration on long voyages, if undiscovered beforehand.

The acoustic TWS system uses discrete narrowband frequencies to detect phase changes in waves reflected from a moving person. By interfering successive return pulses, small changes in phase and amplitude within the reflected beam lead to big changes in voltage waveforms, allowing detection of mm-scale motion by cm-scale wavelengths. This very sensitive but simple means of detecting the mm-scale motion of persons who are breathing but otherwise stationary works *only* for narrowband frequencies. The interference effects are washed out for wideband beams, like those of impulse radars, which are the most common radar TWS systems. These ultra-wideband (UWB) radars generally have a lower bound on detectable velocity of about 12 to 15 cm/s.

The purpose of this section is to explain how an acoustic TWS sensor can be made to detect and locate humans in near real time and can have S/N improved up to about 30 dB through a well-designed filter/amplifier.

Owing to the processing algorithms used, jitter in the start of each pulse of a mechanical transmitter is inconsequential. What we have found to be critical to processing, instead, is that the start time of each pulse be known with an uncertainty, $\Delta t$, less than about 1/20 of a wave period, or ideally no more than one sample time. The means of achieving such a low $\Delta t$ with mechanical transmitters, used in demonstrating the breadboard TWS system in Fig. 4, is the following.

In Fig. 4, one sees leads from two piezo-film sensors that are affixed to the underside of the transmitter plate. When the impactor strikes the transmitter plate, the piezo-film sensors instantaneously produce a sharp voltage spike of tens of volts that establishes the fiducial $t = 0$ with extremely low uncertainty. (One can choose the location of a piezo-film sensor on the transmitter plate to produce a fiducial voltage spike of any lesser voltage, as desired.) In our breadboard demonstration, the voltage spike, unamplified of course, was used to trigger the receiver. The shot-to-shot reproducibility of the start time determination by this method was as accurate as could be measured, that is, within one sample time.

The signal processing for each transmitted pulse begins with the impactor striking the transmitter plate. That event triggers by any means the start of a delay period of some number of milliseconds. The delay period is needed so that the sound pulse produced by the impact and conducted through the air and through the walls to the receivers will be sufficiently diminished at the receivers by the end of the delay period. The downside of the delay period is that persons in the immediate vicinity of the transmitter will not be detected. And the longer the delay period, the greater the 'dead zone' for detection.

After the delay period ends, the signal processor begins acquiring voltage waveforms from the sensors on the receivers. If it is only required to detect humans, and not to locate or track them, then only one receiver may be necessary. Otherwise, a horizontal linear array of at least two receivers is needed. The digitized voltage waveforms from each receiver carry information about the acoustic waves reflected from all objects and persons on the other side of the wall. (Since the receivers may also receive reflections from the same side of the wall, it may be necessary to operate the system remotely if transmitter back lobes and side lobes cannot be suppressed or shielded.) The next steps in signal processing are to remove reflections from completely motionless objects.

At all times, a running average is kept in memory of the past $n$ waveforms from each receiver. Since the noise level will be suppressed and S/N will be increased by a factor of about $n^{1/2}$, the number of waveforms in the average should be as large as allowed by operational constraints. The newest waveform should start at exactly the same time delay with respect to the impactor striking the plate as the average waveform. Or at least the difference in time delays with the average waveform should be very much less than the wave period. If not, then it may be necessary to adjust the start of the newest waveform to correspond to the start of the average waveform. This can be done by cross-correlating the newest waveform with the average waveform.

After $t = 0$ has been established for the newest voltage waveform for each receiver, a narrow bandpass filter is applied to the waveforms. The filter should be matched to the resonant frequency or frequencies of the transmitter and receiver. (The resonances should be identical in both.) And the bandwidth of the filter should be matched to be no less than the full width at half maximum (FWHM) of the spectral resonances. As long as the bandwidth encompasses the FWHM, the narrower the bandpass filter, the more noise is excluded from the signal and the higher the S/N. S/N increases at the margin almost in inverse proportion to the bandwidth. But the bandpass filter should not be so narrow that it excludes a significant part of the in-band resonant energy, or that it reduces probability of detection, as discussed later in this section.

After the waveforms have been filtered, the filtered waveform from each receiver is compared to, and subtracted from, a running average of the past $n$ filtered waveforms from that receiver. That is, the newest filtered waveform is destructively interfered with the most recent average filtered waveform.

The output of that waveform interference process, the difference waveform, should only differ from noise for those reflections from targets that are not motionless. The difference waveform is amplified with an automatic gain control (AGC). The AGC is designed to compensate for the weaker reflections arriving from more distant targets. The power of the diffuse sound waves reflected from a human target scales with range $r$ as about[5] $r^{-3.8}$. The time-of-flight of the reflected sound waves is measured, and the range is half the time-of-flight times the sound speed.

As an example of the application of these signal processing techniques, this section presents actual waveforms produced and measured by the breadboard impact transducer and matched receiver system shown in Fig. 4. The transmitter and receiver plates were identical, and their resonance at $f_0 = 2186$ Hz was used. The quality factor was $Q = 343$, suggesting that the full width at half-maximum (FWHM) intensity of the presumed line shape was about 7 Hz, for a fractional bandwidth of 0.003. The power spectral density (PSD) of the transmitter or receiver plate is shown in Fig. 5. A COTS bandpass filter of 1 to 10 kHz was applied to the PSD in Fig. 5.

The signal processing for the breadboard TWS system did not have a narrowband filter/amplifier or the capability to produce difference waveforms in real time. The results that follow were derived in post-processing. Without a filter/amplifier that would have increased S/N by about 26 dB, as discussed below, the system could only detect humans between two 'walls' along a one-way path from the transmitter to the receiver. The two 'walls' were a massive mahogany desk, which acted as a soundboard (see Sec. 4), and a solid hardwood table, nearly 1-in thick and over 5 m away.

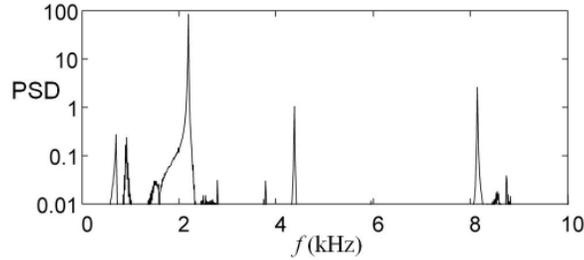

Figure 5. PSD (arb. units) vs. frequency (kHz) for strike test on receiver plate assembly.

Figure 6 shows a comparison in post-processing of three successive through-wall tests, in which the received waveforms were each averaged over 16 pulses by a recording oscilloscope. First, a baseline voltage waveform (fine black in Figs. 6(a) and 6(b)) was created with the operator motionless. Then with the operator position unchanged, a second waveform was created and overlaid in post-processing as the light-gray waveform in Fig. 6(a), covering most of the baseline waveform. Lastly, a third waveform was created as the operator moved about between the two 'walls', and was overlaid as the gray waveform in Fig. 6(b). The heavy black waveforms in both Figs. 6(a) and 6(b) are the differences between the gray and fine black waveforms.

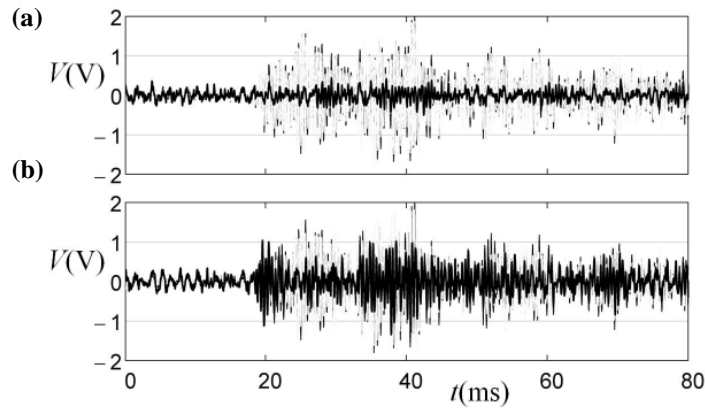

Figure 6. Voltage (V) vs. time (ms) for baseline waveform (fine black) and difference waveform (heavy black) with waveforms (gray) of: (a) stationary person; and (b) moving person.

With the transmitter and receiver about 18 feet apart, after the transmitter triggered the receiver, there was a delay of over 15 ms before the acoustic waves from the transmitter accumulated at the receiver to produce a signal distinguishable from noise, as seen in Figs. 6. In Fig. 6(a), the difference waveform (heavy black) is almost the same as the noise level before the acoustic waves are noticeable at the receiver. In Fig. 5.2(b), the effects of the changed position of the person, evident as a much larger difference waveform (heavy black), are seen throughout the pulse owing to multipath propagation of sound in a cluttered environment.

The PSDs of the received signals do not distinguish between moving and stationary persons as well as the difference waveforms do, because the frequency content of the received signals is not changed nearly as much as their phase. The differences in the frequency content do become significant, however, when comparing the inverse fast Fourier transforms (IFFTs) in a filtered band about the resonance.

Figure 7 compares the IFFTs in filtered bands about the resonant frequency. A numerical bandpass filter was emulated by only performing the IFFT on the Fourier frequency components within a specified frequency interval. In Fig. 7, the IFFT was performed on the 11 Fourier components from 2101 Hz to 2223 Hz, which span a filter bandwidth that is 5.6% of the central frequency of 2162 Hz. The purpose of a narrowband filter/amplifier is to exclude all noise outside the narrow bandpass filter, but still admit most of the useful signal, and thereby increase the S/N. The COTS filter/amplifier used in this breadboard demonstration had a bandpass filter from 1 kHz to 10 kHz. That means it admitted all noise over a 9 kHz range. But, as seen in Fig. 7 and other emulations[26], most of the useful signal is contained only within about a 24 Hz to 120 Hz range. In this example a properly designed filter/amplifier, therefore, could eliminate of the order of about 19 dB to 26 dB of noise from the filtered signal, improving the S/N by like amounts.

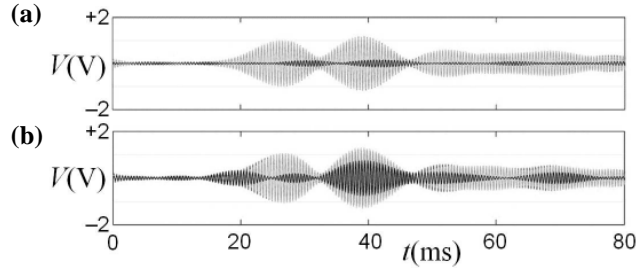

Figure 7. IFFT voltage (V) with 5.6% bandpass filter vs. time (ms) for baseline waveform (fine black) and difference waveform (heavy black) with waveforms (gray) of: (a) stationary person; and (b) moving person.

If the bandpass filter is too narrow, the distinction between a moving person and a stationary person fades. The distinction with a 1.1% bandpass filter cannot be made with as much confidence as the distinction with a 5.6% bandpass filter[26] in Fig. 7. The optimal bandwidth of the bandpass filter, therefore, is a tradeoff between elimination of noise with narrow filters and confidence of detection with wider filters. The bandpass filter should be wide enough to encompass more than a few Fourier components of the FFT. In practical terms, that condition is equivalent to the condition that the bandpass filter should be much wider than the inverse of the record length.

For operational reasons, a prf of a TWS sensor may need to be at least about 3 Hz, which would limit the record length to a few hundred ms. Then the bandpass filter should be much wider than a few Hz, say, at least a few tens of Hz. For the breadboard TWS system, the record length in Figs. 6 and 7 was 81.92 ms, suggesting that a bandpass filter for the breadboard should be at least about 100 Hz wide, or about 5% of the operating frequency of 2.2 kHz. This condition was confirmed by emulations[26], which showed a loss of detection capability with a 1.1% bandpass filter.

Filters that are too narrow should also be avoided because the resonant frequencies can change or drift, depending on many conditions. The resonant frequency of the breadboard receiver, when it is held in the hand and struck like a tuning fork, is 2186 Hz. The resonant frequency when it is fastened by mounting tape onto a wall is 2198 Hz, a difference of 0.5%. To determine the optimal bandwidth and central frequency of a bandpass filter, the system should be tested under a wide range of operating conditions, varying: Wall types; separation of transmitter and receiver; mounting of the transmitter and of the receiver; transmitter/wall air gap; impact power of the impactor on the plate (separation of plate from impactor); range to target (higher frequencies attenuate faster); and any other factors that might affect the spectral response of the receiver.

## 6. BREADBOARD IMPACT TRANSDUCER FOR TWS

This section describes the design and characterizes the performance of the breadboard TWS transducer system, shown in Fig. 4, that was used to demonstrate the detection (with only rudimentary signal processing, in post-processing) of a person moving between two thick solid walls over 5 m apart, as seen in Figs. 6 and 7.

When the center of the face of the receiver plate was tapped lightly with a metal object, the piezoelectric detectors on the plate produce the voltage waveform shown in Fig. 8. In Fig. 8, the best fit of the envelope of the amplitude of the voltage oscillations having the form $V = A\exp(-at)$ is for $A = 3.3$ V and $a = 20$ Hz. Since the quality factor $Q$ is given in terms of the decaying voltage waveform by $Q = \pi f_0 / a$, where $f_0 = 2186$ Hz is the resonant frequency of the receiver plate, we find $Q = 343$.

This estimate of $Q$ is supported by an estimate of the line width of the resonance. Figure 5 showed the PSD of the voltage waveform in Fig. 8. Since the sampling rate in these measurements was 25 kHz and the record length of the waveform for the FFT was 2048 samples, the frequency resolution of the PSD was only 12.2 Hz. Zooming in on the three frequencies about the resonance at 2186 Hz in Fig. 5, however, suggests that the FWHM intensity of the presumed line shape was about 7 Hz, for a fractional bandwidth of 0.003, consistent with the measurement of $1/Q = 0.0029$ above.

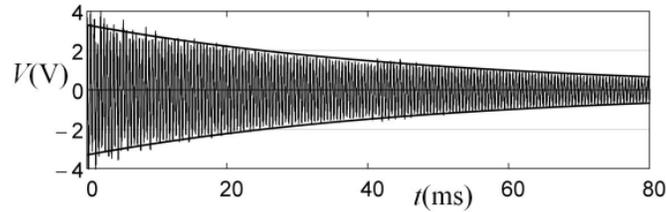

Figure 8. Voltage (V) vs. time (ms) for strike test on receiver plate assembly. Curves are best fit of exponential decay to waveform envelope.

The breadboard impact transmitter works as follows. The actuator for the impactor is a spring pullback and release mechanism, shown in Fig. 9. The steel impactor is welded to its post and to the top of the spring via a cylindrical mounting inside the spring, as seen in Fig. 9(b). The post is pulled back against the spring by a cam follower that follows the spiral arc of the rotating cam seen in Fig. 9(a).

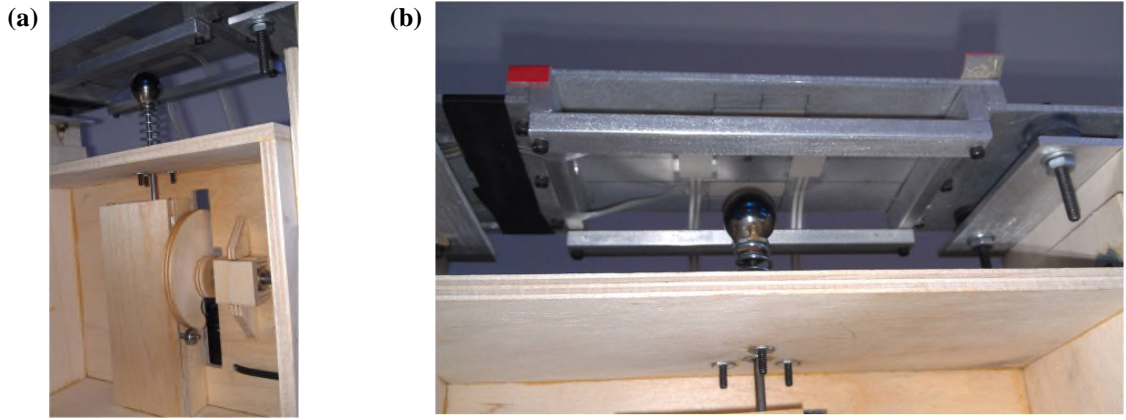

Figure 9. Transmitter: (a) Cam follower has spring almost fully retracted; (b) close-up view of impactor at rear face of transmitter plate with triggering piezo-film sensors affixed.

When the cam follower has compressed the spring to its maximum amplitude, the cam follower falls over the lip of the cutout in the cam, seen in Fig. 9(a). The sudden release of the cam follower by the cam releases the post to which the impactor is welded, and the compressed spring is 'unlatched' to drive the impactor against the rear face of the transmitter plate, seen in Fig. 9(b). The cycle then repeats, with the outer edge of the rotating cam smoothly picking up the cam follower at small radius and pushing the cam follower against the increasing force of the spring. The rotating cam is driven by the motor and starter capacitor seen in Fig. 4. In case the motor is switched on when the spring is nearly fully compressed by the cam follower, the starter capacitor ensures a smooth startup.

The transmitter plate assembly is identical in all respects to the receiver plate assembly, seen in Fig. 10. In fact, the two assemblies are interchangeable, except that the leads from the piezo-film sensors affixed to the receiver plate are soldered together in series. The transmitter plate assembly is attached to the body of the transmitter by means of bolts and rubber acoustic isolators. The height of the transmitter plate assembly above the impactor is adjustable to optimize performance.

The L-shaped cutouts at the corners of the transmitter-plate and receiver-plate assemblies, seen in Fig. 10, allow the position of the transmitter plate to be shifted along both axes parallel to the plate. This capability allows higher-frequency modes of the transmitter plate to be excited by off-center impacts. With four piezo-film sensors arrayed around the center of the plate, we were able to compare the voltage waveforms in each and spectrally analyze the plate resonances one-by-one to determine the nature of the plate normal modes at each resonant frequency.

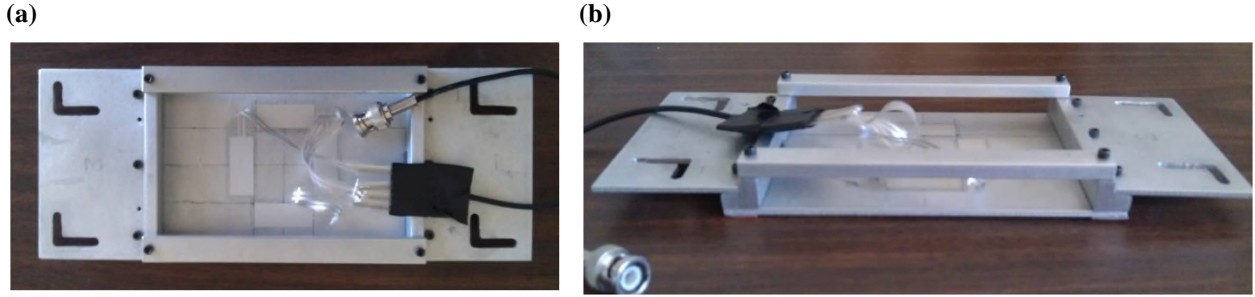

Figure 10. Receiver plate: (a) Rear face showing four piezo-film sensors connected in series at a 10-pin header (covered with tape); (b) mounted on wall with 1-mm plate-wall gap.

The sound pressure level (SPL) of this impact transmitter was not measured. The SPL of an earlier, less powerful version of the impact transmitter, however, was measured. The calibrated root-mean-square (rms) pressure measured during the first 4 ms of the transmitter pulse was $P_{rms}$ = 9.02 Pa. During that 4-ms interval, therefore, the transmitter SPL was 113 dB *re* 20 μPa at 1 m. This SPL is just 7 dB below the threshold of feeling in the ear[28]. At this SPL, the suggested daily noise exposure level for nonoccupational noise is less than 2 minutes[28]. Indeed, to avoid discomfort, it has been necessary to wear ear protection when working with either transmitter, particularly since we were sometimes closer to the transmitter than 1 m. The in-band rms pressure in a 12.5% bandwidth during the first 4 ms of the transmitter pulse was measured to be $P_{rms}$ = 7.47 Pa. During that 4-ms interval, therefore, the in-band SPL in a 12.5% bandwidth was 111 dB *re* 20 μPa at 1 m.

## 7. CONCEPTUAL DESIGN OF ACOUSTIC STOWAWAY DETECTION SYSTEM

Embodied in a product, the TWS technology presented in this paper offers the potential for rapid nonintrusive detection of stowaways inside closed steel cargo containers, truck bodies, and train cars. Inducing an acoustic pulse at any location on the outer shell of a container creates an acoustic wave pattern in the air inside the container. If the induced pulses are identical, pulse-to-pulse, then the wave patterns created in the air inside the container will also be identical, pulse-to-pulse, if and only if there is no motion inside the container from one pulse to the next.

The method of detection is to store the received signal, usually averaged over some number of pulses to increase signal-to-noise, and then interfere the received signal with subsequent pulses. Even a slight difference in phase of the subsequent pulses results in a large difference in the received signal when the signals are interfered by subtraction.

Typical steel sea cargo containers range in length from about 10 ft to 48 ft, with 40 ft (12 m) perhaps being the most common. Typical cross-sectional dimensions are from about 8 ft to 9.5 ft. An average 40-ft shipping container weighs about 10,000 pounds. Because cargo containers may be packed closely, access to each container may be limited to an end or a side only. In any event, high throughputs are more readily achieved by inspecting the containers from one access point only.

The TWS system may be sensitive to motion on the operator's side of the wall. (The acoustic vibrations in the wall get transmitted into air in both directions.) This effect should not be significant for cargo containers with a transmitter at one end and a receiver at the other. This effect is mitigated either by shielding the transmitter from the operator or by operating the transmitter remotely. But remote operation may be desirable anyway for expeditious high-throughput processing of cargo containers.

Figure 11 schematically illustrates a remotely operated TWS system that accesses each container at only one point. It has telescoping arms, vertically and horizontally, to access containers that are stacked irregularly. The resonant-plate transducer is mounted on the end of the horizontal arm.

The choice of operating frequency is a trade-off between wavelengths short enough to be sensitive to small motions and long enough to be able to produce high acoustic powers easily. The method of sensing motion by interfering successive waveforms results in a sensitivity to displacements of the order of about one-tenth of a wavelength. That is, to be sensitive to mm-scale displacements, characteristic of a single person lying down and breathing quietly, the wavelength should be not longer than about 1 cm, and the frequency should be not less than about 30 kHz. To be sensitive to cm-scale displacements of a single person, the frequency should be not less than about 3 kHz. It is much easier to produce high powers at 3 kHz than at 30 kHz, and that is the trade-off.

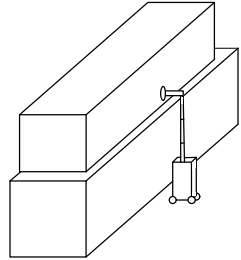

Figure 11. Schematic illustration of remotely operated TWS for inspecting cargo containers.

If a cargo container contains persons hidden among inanimate objects, such as typical cargo, then detection of the persons is possible, but more challenging. The container is a lossy cavity in which the acoustic wave pattern of a single tap changes and decays with time. The method of detecting persons is sensitive to small changes from pulse to pulse of the acoustic wave pattern inside the container. By diffraction, the wave pattern more or less fills all the airspace within the container. As long as the diffracted wave pattern has sufficient amplitude at the location of the person, relative to the amplitude elsewhere in the container, the person should be detectable. A displacement of the person must lead to a detectable perturbation of the phase of the wave (by at least about one-tenth of a wavelength) at the location of the sensor. This requirement drives the trade-off of frequencies to longer wavelengths, which diffract more.

The pulse repetition frequency (prf) is limited by the time it takes for the acoustic wave pattern inside the container to decay away sufficiently, such that there is no cumulative build up of residual amplitudes over a succession of pulses. In a 40-ft (12-m) cargo container, sound traveling through the steel walls takes about 2 ms to travel from one end to the other. Through air, sound takes about 70 ms to travel round-trip, from end to end and back. These considerations suggest the prf should perhaps be not more than about a few Hz. To improve the signal-to-noise ratio, the received signal should be averaged over about 10 pulses, which is shorter than the time to complete one breathing cycle. If multiple persons are concealed within a container, then a positive detection should occur in less than a few seconds.

Probability of detection and false alarm rate depend strongly on (a) the threat scenario, (b) construction of the container, and (c) choice of operating parameters. The threat scenario includes number of persons in a container, their state of motion, and whether and how they might be concealed among inanimate cargo. The construction of the container includes the material and thickness of the walls as well as the sturdiness and rigidity of the construction. Introducing an acoustic wave into the interior requires causing the walls physically to vibrate. Much higher acoustic powers can be transmitted through thin, flexible walls than thick, rigid walls with ribbed construction. The choice of operating parameters includes frequency, bandwidth, and power. This choice depends on the supposition of threat scenario. One might choose to design the system such that the likelier scenarios have higher probability of detection.

The breadboard TWS system demonstration described in Sec. 6 had only rudimentary signal processing capabilities and required post-processing for human detection. The following discusses how the breadboard signal processing would need to be augmented for actual TWS applications, like scanning cargo containers for detection of stowaways, to provide near-real-time signal processing capability and an increase of S/N by several tens of dB.

Because the operating frequency of the breadboard transducer system, less than about 2.2 kHz, is relatively low, the sampling rate requirements are modest. In order to accommodate round-trip travel times of acoustic waves in cargo containers, most of which are 40-ft (12-m) long, especially when the containers might be cluttered and the acoustic waves might take some time to reverberate and accumulate at the receiver, the record length should be at least about 80 ms and need not be more than about 160 ms. According to the Nyquist criterion, the lowest sampling rate that could possibly measure the resonance is about 5 kS/s, but such a low rate would lead to aliasing errors. Better is a sampling rate of 10 kS/s, and more than adequate is a sampling rate of 25 kS/s.

The signal processor for a TWS product should therefore be designed to accept the signals from a narrowband filter/amplifier at a pulse repetition frequency (prf) of at least 3 Hz. (This prf is about the quickest that allows the reverberations from the previous pulse to decay.) Following are the signal-processing steps that will need to be performed with the waveform output from the bandpass filter:
(i) The $t = 0$ mark on the waveform should be established by the methods described above.
(ii) The voltage waveform should be centered about the mean voltage.
(iii) The voltage waveform should be stored in memory.
(iv) A running average of the past $n$ waveforms, where $n$ is 10 to 16, should be constantly updated in memory. The new $(i+1)^{th}$ waveform should be subtracted from the running average of the past $n$ waveforms, up to and including the $i^{th}$ waveform, in order to produce a difference waveform.
(v) If this difference waveform exceeds a threshold voltage over multiple contiguous samples, that indicates a detection.
(vi) Every time the new $(i+1)^{th}$ waveform is added to the average, the $(i+1-n)^{th}$ waveform should be removed from the running average of the past $n$ waveforms to update the running average.
(vii) At a prf of 3 Hz, the first real-time detection can be made within about 3 to 5 seconds.

This method of real-time detection by exceedances of a threshold voltage is illustrated in Fig. 12. The data from the baseline, second, and third waveforms shown in Fig. 6 were used with a thresholding algorithm that only displays voltage exceedances above 0.5 V. Since the noise amplitude is almost entirely below 0.5 V, voltage differences above this threshold caused by human motion are readily apparent. (The tapering-off of exceedances at late times would not occur with an AGC.)

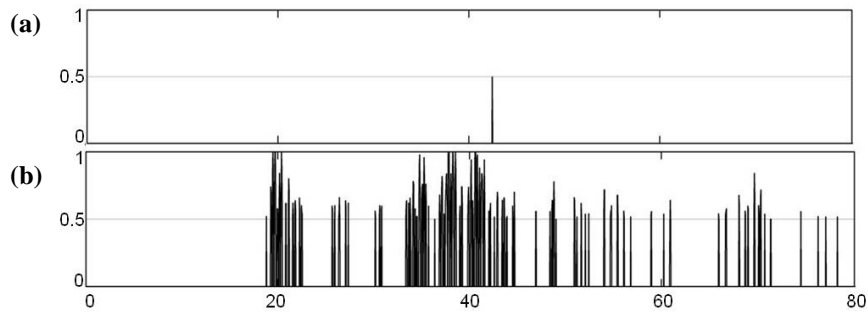

Figure 12. Demonstration of through-wall detection based on difference waveform data from Fig. 6. Threshold-detection display (volts *vs*. time in ms) for: (a) stationary person; (b) moving person.

Noteworthy about Fig. 12 is that the detection of a moving person was accomplished without a narrow bandpass filter or FFTs or IFFTs. The detection involved just a subtraction of waveforms and amplification of the difference waveform without even an AGC, suggesting that just rudimentary signal processing may be adequate for TWS with a sufficiently powerful transmitter, such as the impact transmitter used here. If tracking is needed, however, and not just detection, then some of the signal processing methods discussed in Ref. [26], and particularly AGC, may be necessary for the tracking algorithms to work well.

(Feb. 1999), <https://www.ncjrs.gov/pdffiles1/nij/grants/178564.pdf>.